# Complexity synchronization analysis of neurophysiological data: Theory and methods

Schizas, Ioannis[a,1], Sullivan, Sabrina[a,1], Kerick, Scott E.[a], Mahmoodi, Korosh[a,*], Bradford, J. Cortney[a], Boothe, David L.[a], Franaszczuk, Piotr J.[a,b], Grigolini, Paolo[c], West, Bruce J.[c,d]

[a]US Army Combat Capabilities Development Command, Army Research Laboratory, Aberdeen Proving Ground, MD, USA
[b]Department of Neurology, Johns Hopkins University School of Medicine, Baltimore, MD, USA
[c]Center for Nonlinear Science, University of North Texas, Denton, TX, USA
[d]Office of Research and Innovation, North Carolina State University, Raleigh, NC, USA

**Abstract**

We present a theoretical foundation based on the spontaneous self-organized temporal criticality (SOTC) and multifractal dimensionality to model complex neurophysiological and behavioral systems to infer the optimal empirical transfer of information among them. We hypothesize that heterogeneous time series characterizing brain, heart, and lung organ-networks (ONs) are necessarily multifractal, whose level of complexity and, therefore, its information content is measured by its multifractal dimension. We apply modified diffusion entropy analysis (MDEA) to assess multifractal dimensions of ON time series (ONTS), and complexity synchronization (CS) analysis to infer information transfer among ONs that are part of a network-of-organ-networks (NoONs). The purpose of this paper is to advance the validation, standardization, and repeatability of MDEA and CS analysis of heterogeneous neurophysiological time series data. Results from processing these datasets show that the complexity of brain, heart, and lung ONTS significantly co-vary over time during cognitive task performance but that certain principles, guidelines, and strategies for the application of MDEA analysis need consideration. We conclude with a summary of the MDEA's limitations and future research directions.

## 1. Introduction

Dynamic interactions among brain, heart and lung organ networks (ONs) may be considered a co-evolution of information exchange among multilayer, multifractal integrated ONs Bartsch et al. (2015); Ivanov (2021); Mahmoodi et al. (2023a); Marzbanrad et al. (2020); West et al. (2023b). Such co-evolution or complexity synchronization (CS) enables flexibility and adaptability of the human organism as well as robustness at the interface of ever changing internal and external environment demands and contexts. New theories and data processing methods are needed to better understand interactions among complex systems (NoONs) and to advance human-human and human-machine interaction technologies and interventions (e.g., see West et al. (2023a)).

Here, we base our analysis of neurophysiological ON-interactions on the theory of multifractal dimensionality and crucial events (CEs) emerging from spontaneous self-organized temporal criticality (SOTC) Mahmoodi et al. (2023a); West et al. (2023b). SOTC is a bottom-up process of cooperative interaction of components of a complex system (a NoON in the present context) by which spontaneous behavior of the whole emerges, and this research provides a conceptual/analytical framework for investigating such principles within complex systems (NoONs). CS is characterized by high-order synchrony among the varying inverse power law (IPL) scaling indices ($\delta's$) of interacting complex systems (ONs), which we hypothesize is the mechanism necessary for coordination among them Mahmoodi et al.

---

*Corresponding author. Email: koroshmahmoodi@gmail.com
[1]These authors contributed equally to this work and share first authorship.

(2023a); West et al. (2023b) and we posit that CS is a foundational principle underlying how information is transmitted within and among complex neurophysiological ONs within and among individuals over various timescales.

Recently, we applied modified diffusion entropy analysis (MDEA) to electroencephalographic (EEG), electrocardiographic (ECG), and respiratory (RESP) time series data simultaneously recorded during cognitive task performance Mahmoodi et al. (2023a); West et al. (2023b). Preliminary results showed that using this approach we observed synchronization of complexity scaling indices ($\delta's$) across 64 channels of EEG, along with single channels of ECG and RESP, despite the drastic differences in the temporal dynamics and frequency scales of these three heterogeneous ON time series. However, analyses have been limited to data from only two participants during the performance of two different tasks (neurofeedback training and the Go-NoGo task) as our preliminary proof of concept. In attempting to apply these analyses to data from the full dataset (and other datasets) and relate the observed CS to behavioral performance, we acknowledge that the preliminary analyses have been overly simplified and that various interactions among numerous factors and parameters need to be taken into consideration for the method to be generalizable and repeatable by independent researchers across diverse studies and multimodal datasets.

For example, experimental design, individual differences of subjects, task type and structure, recording duration, sampling rates of diverse time series data (biological, neural, physiological, behavioral), types and levels of preprocessing or decompositions of the different data types, artifact considerations, and missing data or discontinuities in recordings. Further, theoretical development and systematic testing of the MDEA algorithm (e.g., parameter tun- ing for the number of stripes, IPL fit region, and fit method) are required so that common principles and practical guidelines can be implemented to enable repeatability and generalizability of CS analyses.

Herein, we present principles and guidelines for CS analysis and report systematic testing and further development of the analyses based on MDEA combined with automated parameter selection applied to simulated as well as to empirical data (EEG, ECG, and RESP). Thus, the stated purpose of this paper is to advance the validation, standardization, and repeatability of MDEA for CS analysis. We also provide data and Matlab code to facilitate further refinements and to promote future research progress Github (2024).

*Experiment design and task*

Experiments are designed and specific tasks or paradigms are implemented to address specific research questions and hypotheses. Often, when testing new theories and analytical approaches, experiments have yet to be designed and conducted, while analyses can be applied to existing and simulated data in a preliminary stage to gain new insights and help better formulate there yet-to-be-done experiments. However, because the use of existing data comes with the limitation that the data do not originate from an experiment designed for the purposes of testing the current theories and analyses, there will invariably be challenges and limitations that must be taken into consideration. For example, in the present work, we are interested in communication among neural, physiological, and behavioral processes that interact in complex dynamic ways during cognitive task performance. We selected data from a recent neurofeedback study Kerick et al. (2023) because multiple sources of data (EEG, ECG, RESP, behavior) were simultaneously recorded from multiple subjects ($N = 30$) during cognitive task performance in low and high time stress conditions (Go- NoGo task). Limitations of using these data are that the nature of the task, the task structure, and recording duration and comparative conditions were designed with more traditional methods of analysis in mind, so may not be best- suited to answer research questions based on complexity theory. However, the advantages are that the data consist of simultaneously recorded neural, cardio-respiratory, and behavioral time series, which enables the leveraging of these data to test hypotheses regarding how heterogeneous complex systems (NoONs) interact, and how such interactions relate to behavioral task performance.

*Data considerations*

Neurophysiological and behavioral data manifest stochastic and deterministic properties with varying degrees of stationary, quasi-stationary, and non-stationary segments and often exhibit periodic, aperiodic, and intermittent dynamics (see Figures 1 and 2). It would seem that MDEA processed signals with recurring (ECG) or highly periodic (RESP) patterns that would identify differing sequences of independent events than with more stochastic signals (EEG). Events are defined as the transition of the time series across amplitude thresholds (see section MDEA below for more details). Events detected from such signals as ECG and RESP would likely result in highly correlated inter-event intervals generated from the analysis, whereas those from EEG would likely exhibit less correlation among events. This has yet to be empirically determined with real and simulated data, which we investigate here. Such features of heterogeneous time series present challenges for most methods of coupling or synchronization for signals of the same type (e.g., multi-channel EEG data) or of different types (e.g., brain-heart).

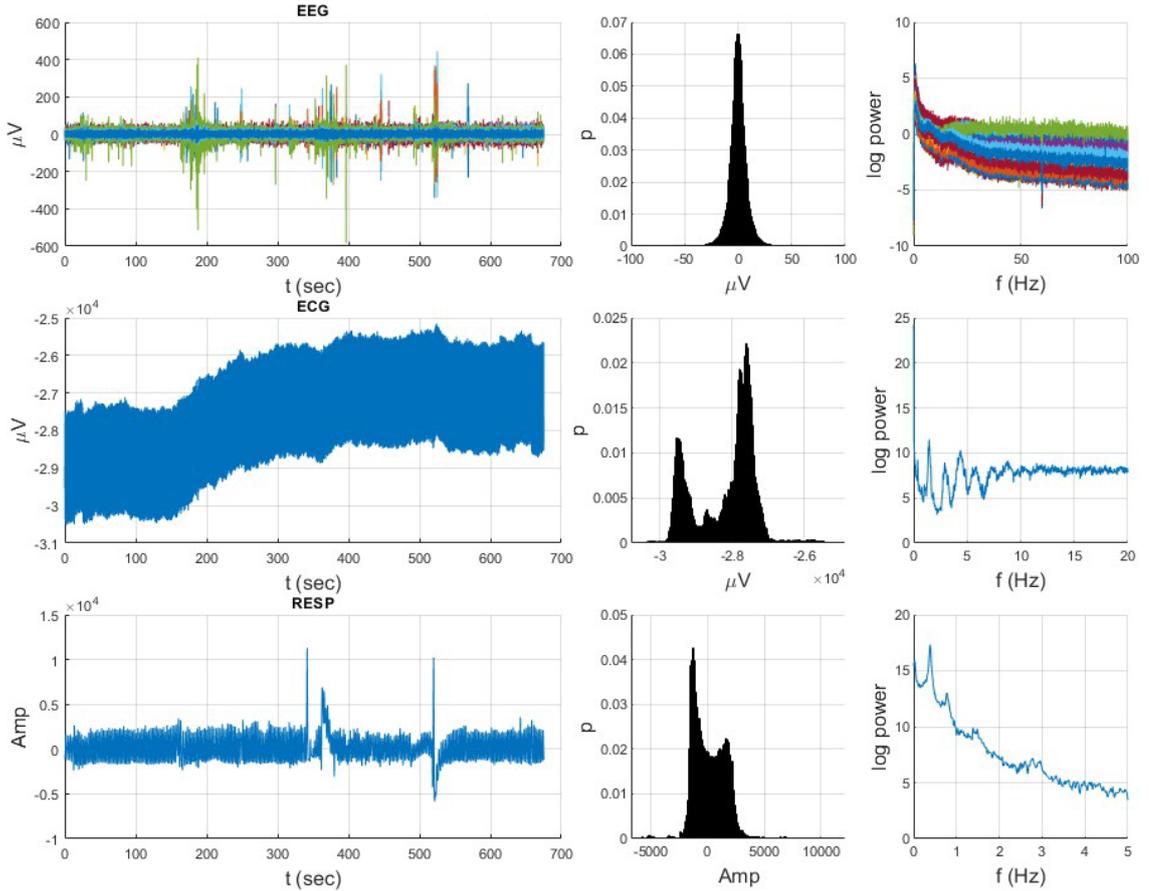

Figure 1: Heterogeneous neurophysiological data from an example subject over the 675-s duration of a Go-NoGo task (64-chan EEG, ECG, RESP) illustrating diverse temporal dynamics, distributions, and spectra.

*1.1. Sampling rates, data length, recording/window duration, and pre-processing*

Sampling rates of time series data are governed by the Nyquist theorem Nyquist (1928), which states that periodic data must be sampled at twice the rate of the highest frequency component of the signal. Because we may not know the precise highest frequency, in practice, we use low pass filters to limit the frequency band, and we use a sampling rate four or more times the cutoff frequency of the filter to prevent any aliasing. Higher sampling rates may be beneficial for some analyses where time resolution is important but also may be detrimental for other analyses (e.g., auto-regressive) due to longer intervals of correlated samples. Depending on the particular signal and recording method there might be different optimal sampling rates. This is especially true for EEG data because higher frequencies (e.g., > 100 Hz) are typically less-well studied, and their functional relevance is largely unknown. On the other hand, oversampling may introduce high frequency noise of unknown origin into the time series, for example, measurement (recording system), biological (muscle), or environmental noise (60 Hz), especially for ECG (0-10 Hz) and RESP (0-.5 Hz) time series data which function on much slower timescales than EEG. Further complicating the matter, ECG time series, although exhibiting recurring patterns (P, QRS, and T complexes), are not periodic in the sense of oscillatory signals. Whereas the RESP time series is highly periodic, the EEG time series is predominantly quasiperiodic and aperiodic. Thus, the question is how do we investigate interactions among these vastly different time series? Traditional analyses based on correlation, coherence, phase lags, and other common time series analyses (assuming independence, normality, and stationarity) are not well-suited for investigating complex nonlinear interactions among heterogeneous time series such as EEG, ECG, and RESP.

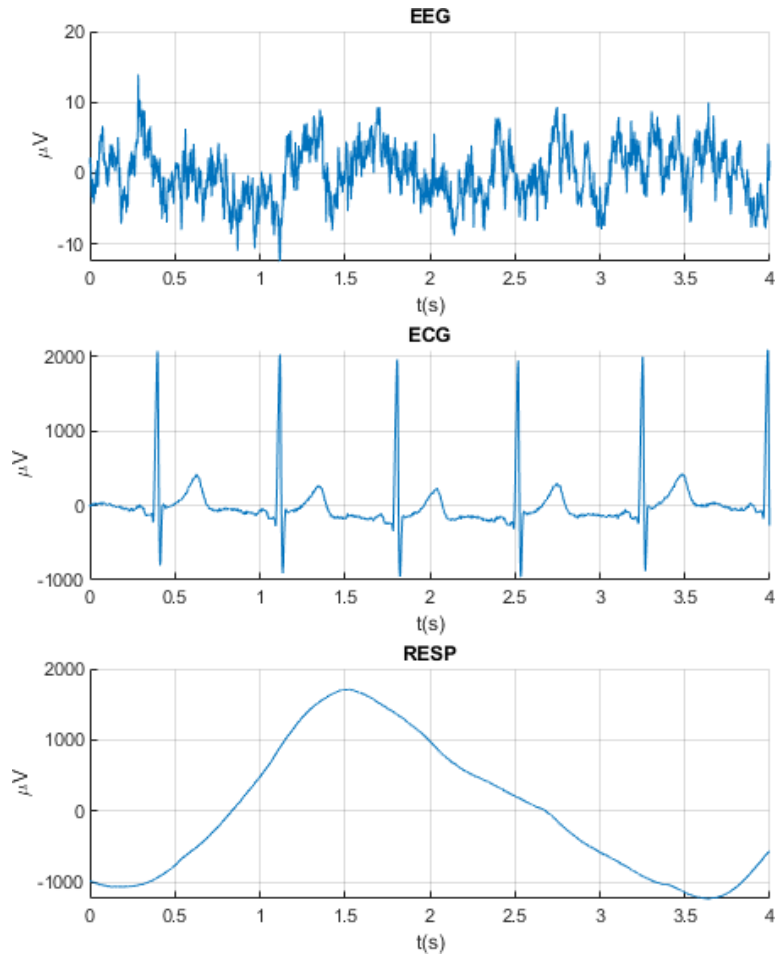

Figure 2: Amplitudes and temporal dynamics of EEG, ECG, and RESP within one breath cycle (4 s), showing the need of having a method to capture the different relevant features of each time series. In MDEA analysis, amplitudes of all time series are normalized to the interval [0, 1], but the original amplitude and frequency scales are shown here to illustrate the different orders of magnitude scales of EEG with respect to ECG and RESP.

Table 1 records some of the interesting and more advanced approaches implemented in the literature for the investigation of coupling between the brain and the heart and/or the heart and lungs. We do not go into an in-depth review of these methods here, only to say that we believe our novel theory-driven approach based on CS derived from MDEA adds a significant contribution to this area of research. Future research would profit greatly by focusing on comparisons of various methods aimed at better understanding interactions among two or more complex systems

(NoONs).

| |
|---|
| **Linear** |
| Time-Delay Stability Bartsch et al. (2015); Bartsch and Ivanov (2014); Liu et al. (2015) |
| Controlled Time Delay Stability Alskafi et al. (2023); Marzbanrad et al. (2020) |
| Delay Correlation Landscape Lin et al. (2016) |
| Time-Variant Coherence Piper et al. (2014) |
| Cross Spectrum Analysis Herrero et al. (2018) |
| Partial Directed Coherence Leistritz et al. (2013) |
| Phase Synchronization Analysis Bartsch and Ivanov (2014); Rosenblum et al. (1996) |
| Phase-Amplitude Coupling / Comodulation Maps Tort et al. (2010, 2018); Canolty and Knight (2010) |
| Multivariate Interaction Analysis Pernice et al. (2019) |
| Heartbeat-Evoked Potentials Petzschner et al. (2019); Schandry et al. (1986) |
| **Nonlinear** |
| Convergent Cross Mapping Schiecke et al. (2019); Sugihara et al. (2012) |
| Recurrence Quantification Analysis Martin et al. (2015); Marwan et al. (2007) |
| Transfer Entropy Catrambone et al. (2021); Schreiber (2000); Vicente et al. (2011) |
| Directed Transfer Entropy Deco et al. (2021) |
| Phase Transfer Entropy Lobier et al. (2014) |
| Conditional Entropy Kumar et al. (2020) |
| Cross-Sample Entropy Martin et al. (2015) |
| Joint Distribution Entropy Li et al. (2016) |
| Multiscale Entropy Costa et al. (2005); Gao et al. (2015); Jelinek et al. (2021); Pan et al. (2016) |
| Diffusion Entropy Analysis Scafetta and Grigolini (2002) |
| Mutual Information Kotiuchyi et al. (2021) |
| Interaction Information Decomposition / Partial Information Decomposition Faes et al. (2017) |
| Maximal Information Coefficient Catrambone et al. (2021); Reshef et al. (2011) |

Table 1: Methods for Neurophysiological time series coupling analyses

*1.2. Filtering/Resampling*

Many strategies exist to decompose neurophysiological time series data into frequency bands, empirical modes, independent components, principal components, time-frequency atoms, power modes, etc., especially in EEG analysis. For ECG analysis, researchers predominantly study the RR time interval series or heart rate variability (HRV). As such, when asking the question of how complex time series of various origins interact, considerations as to what, if any, transformations or decompositions of the data may be appropriate, and if so, why. Further, various data transformations may render interpretation of the results more difficult and/or more extensive (e.g., analyses conducted across multiple frequency bands or signal components). For these reasons, we opt to preserve the original time series data with minimal pre-processing or decomposing (i.e., high-dimensionality is preserved, avoiding issues associated with decompositions or transformations) for our analyses of CS among EEG, ECG, and RESP. MDEA does not rely on particular oscillatory components, however, prominent oscillatory components may distort measurements of scaling of underlying IPL processes. Here, all data were originally sampled at 2048 Hz, which we then down-sampled to 512 Hz in our previous work Mahmoodi et al. (2023a); West et al. (2023b). EEG data were high-pass filtered at 1 Hz, while ECG and RESP were initially left unfiltered. Independent component analysis (ICA) was also applied to the EEG data to remove eye blinks and saccades, for further details see Kerick et al. (2023); artifacts of various types exist to various extents in most datasets, e.g., transient movement and muscle artifacts. In addition to the above considerations, it is important to also test window lengths used in the MDEA processing and window overlaps on the different time-series data.

*1.3. Artifact-reduction/removal*

Various types and levels of non-brain and non-physiological artifacts are common in studies during recordings while subjects perform various cognitive and behavioral tasks. These various artifacts can be of biological (eye movements and saccades, muscle activity, motion artifacts) or non-biological origin (e.g., 60 Hz line noise, loose electrodes or sensors) and may persist for extended time periods (seconds to minutes) or may be transient (milliseconds to seconds), and they may be localized or global (e.g., a few EEG recording electrodes or all recording electrodes). In event-related paradigms, where data are time-locked to stimuli or responses in short epochs (e.g., milliseconds to seconds), individual trial epochs contaminated by artifacts can be deleted, and then ensemble averaged over trials for analysis. However, for analyses applied to continuous recordings of long duration (minutes to hours), one must decide whether the available signal processing methods are suitable to minimize the artifacts, or whether data need to be cropped or cut from the continuous recordings, thus leaving discontinuities in the remaining dataset. In these cases, appropriate data simulations are necessary to incorporate so that known signals can be superimposed with known perturbations that simulate various types and levels of artifacts and discontinuities observed in empirical data. We are in the process of testing the effects of various EEG artifacts and mitigation strategies on MDEA and CS analysis, but due to the preliminary nature of the testing results, we do not have conclusions to include herein.

*1.4. Missing data/discontinuities*

Because we are interested in the interactions among multiple NoONs over multiple time scales, missing data or data streams interrupted by task breaks or multiple recording sessions that limits the time scales across which the data can be analyzed (see Mahmoodi et al. (2023b) for such application to reaction time series data recorded over multiple sessions separated by several days). In the datasets we analyze herein, we only consider continuous data over approximately 10-12 min task duration in a single session, and missing data and data discontinuities are not encountered.

## 2. Methods

*2.1. Modified diffusion entropy analysis (MDEA)*

Modified Diffusion Entropy Analysis (MDEA) is a processing method devised to detect temporal complexity in time series data Scafetta and Grigolini (2002); Culbreth (2021); Buongiorno Nardelli et al. (2022). The method converts the input time series signal into a diffusion trajectory based on the detection of crucial events (CE). The CE are the transition times of the time series signal from one state to another where the states are defined using stripes, dividing the amplitude of the signal to distinct regions. This is a coarse-graining of the time series signal and creates a binary representation of the signal by 1's for events and 0's otherwise. The accumulative summation of this event time series generates the diffusion trajectory. To make statistics of this single diffusion trajectory, the MDEA method slices it into segments with a moving window. Bringing the beginning of these segments to start from the same origin, the distribution of their endpoint can be measured, and so, for any given window length L, its Shannon-Wiener (SW) entropy can be evaluated. By some simple algebra Scafetta and Grigolini (2002), it can be shown that the SW-entropy is a linear function of the log(L), where the slope $\delta$ is the scaling parameter of the diffusion trajectory. The evaluated scaling $\delta$ is connected to the temporal complexity index $\mu$, which is the power index of the waiting time distribution PDF of the time distances between the two consecutive events $\tau$s. If the waiting time distribution PDF of the $\tau$s has an inverse power law (IPL) form of $\psi(\tau) \propto 1/\tau^{\mu}$, the system has temporal complexity scaled with $\mu$ which is related to scaling $\delta$ by $\mu = 1 + 1/\delta$. For a graphical description of MDEA see Mahmoodi et al. (2023a). MDEA is implemented by functions MDEA.m and MDEA_z.m in Github (2024)

*2.2. Why does MDEA matter?*

For most physiological and behavioral data, the distribution of parameters representing system dynamics is not Gaussian; instead, these distributions often follow an inverse power law (IPL) pattern with long tails. As a result, traditional measures like mean or variance can misrepresent the system's characteristics. For instance, in a given population, a small number of billionaires can dramatically inflate the average wealth, creating a misleading impression of a typical person's wealth. Consequently, it is important to approach average- or variance-based metrics (e.g., detrended fluctuation analysis, DFA) with caution when analyzing complex data in medicine or other fields.

Specifically, when analyzing the temporal behavior of a time series, if the IPL index of the CE waiting-time distribution lies within $1 < \mu < 3$, the system exhibits temporal complexity. For $1 < \mu < 2$, both the mean and variance of the waiting times ($\tau$'s) diverge, while for $2 \leq \mu < 3$ (with $\mu \simeq 2$ in a healthy brain Allegrini et al. (2009)), the mean exists, but the variance diverges. MDEA overcomes these limitations by providing an accurate measure of complexity that does not depend on mean or variance. Additionally, MDEA offers the advantage of assessing the complexity of a single time series, making it a reliable measure for real-time data, such as EEG.

*2.3. How are parameters selected for MDEA?*

As mentioned above, setting stripe sizes is akin to coarse-graining, where broad stripe sizes capture large amplitude variations but may miss smaller fluctuations, and narrow stripe sizes detect smaller fluctuations but may also capture physiologically irrelevant noise. In selecting an optimal stripe size, it is necessary to balance the detection of CEs against minimizing the capture of irrelevant noise. Related to this issue is determining sampling rates and window lengths of the different time series. For example, down-sampling RESP time series from 512 Hz to 4 Hz still accurately represents the oscillatory frequencies of respiration, but it presents an issue for MDEA analysis as it decreases the number of samples in any given window length, thereby negatively affecting the statistics underlying the analysis (e.g., a 30-sec window of data sampled at 512 Hz yields 15360 samples, while the same data sampled over the same duration at 4 Hz yields 120 samples). Hence, a pertinent question is what the optimal data length, sampling rate, and stripe size for MDEA analysis of vastly different time series should be and how do we determine optimal parameter values? How should data comprising orders of magnitude differences in time and frequency scales be analyzed with respect to how they interact over different time and frequency scales (e.g., EEG vs RESP)?

Two crucial issues in the application of MDEA, particularly over large-scale datasets, are: 1) developing a rigorous method for determining how many stripes to implement and 2) determining the linear fitting interval of the diffusion entropy from which the slope of the plot of the SW entropy at a time denoted by window length *w* given by $S(w)$ versus the logarithm of the window time $log(w)$ is used to extract the complexity scaling index $\delta$ Culbreth (2021). This is another way to emphasize the focus of this paper.

*2.3.1. Automated stripe size parameter selection for physiological time-series*

To address the aforementioned questions regarding how to optimize the choice of empirical parameters, we introduce an automated stripe-size selection method building on the property that CE time-intervals should follow an IPL PDF according to the physical model utilized by Mahmoodi et al. (2023a).

$$p(\tau) \propto \tau^{-\mu}, \quad 1 \leq \mu \leq 3, \tag{1}$$

where τ indicates the time-interval between two successive CEs, while μ corresponds to the IPL decay parameter. This further results in a complementary cumulative distribution function (CCDF) of the form

$$F_{pow}(\tau) = P(T > \tau) = \left(\frac{\tau}{\tau_{min}}\right)^{1-\mu}, \tag{2}$$

where $\tau_{min}$ is the smallest possible delay which in our setting is equal to the sampling period, i.e, $\tau_{min} = 1/512$ s (approximately 2 ms) and $T$ is a random variable corresponding to the CE time-interval values.

The stripe size selection can affect the PDF of the time-intervals between CEs. Fig. 3 shows that if the stripe size is not properly selected, then the empirical distribution may not follow an IPL [Fig. 3 (left)], while when properly set (details follow), it results in an empirical distribution closely fitted to an IPL PDF for a physiologically reasonable IPL index μ [Fig. 3 (right)]. Note that the IPL index values depicted are merely indicators of what values could reasonably fit the data (right column) and that no matter what value of the index was selected, the IPL functional form could not fit the data (left column). The left column in Fig. 3 depicts the empirical CCDF for EEG, ECG and RESP signals (from top to bottom), along with theoretical IPL CCDFs for different μ parameters. Similarly, the right column in Fig. 3 depicts the empirical and theoretical IPL CCDF when a proper stripe size is selected. Carefully choosing the stripe size can make sure that the CE time-intervals can closely match an IPL PDF. The question that arises next is how to quantify the goodness of fitting an IPL PDF analytically.

We utilize the Kolmogorov-Smirnov (KS) statistic (Corder and Foreman, 2014) to quantify how well the CE time-intervals follow an IPL PDF for a given stripe size( see function Kolm_Smirn.m in Github (2024)). The KS statistic acts as a fitting measure quantifying how well the time-intervals between CEs, extracted using a given number of stripes, follow an IPL PDF. The KS statistics quantifies the maximum absolute difference between the IPL CCDF of empirical CE time-intervals and a candidate theoretical IPL PDF as a function of the stripe size:

$$D_{N_{samp}}(\mu, s_l) = Sup_\tau \left| F_{emp}(\tau) - \left(\frac{\tau}{\tau_{min}}\right)^{1-\mu} \right| \tag{3}$$

where $s_\ell$ indicates the stripe size while $F_{emp}(\tau)$ corresponds to the empirical CCDF that can be evaluated as

$$F_{emp}(\tau) = \hat{P}(T > \tau) = 1 - \frac{1}{N_{samp}} \sum_{i=1}^{N_{samp}} \mathbf{1}_{\tau_i \leq T}, \tag{4}$$

providing an estimate of the probability that the inter-arrival time variable τ < $T$, with $N_{samp}$ denoting the number of measurements and $\mathbf{1}_{\tau_i \leq T}$ is an indicator function equal to 1 if the $i^{th}$ inter-arrival time realization satisfies $\tau_i \leq T$ and 0 otherwise. Then the optimal IPL parameter μ and stripe size are selected to minimize the KS statistic provided below

$$(\hat{\mu}, \hat{s_l}) = arg\, min_{(\mu, s_l)} D_{N_{samp}}(\mu, s_l) \tag{5}$$

The minimization is performed by conducting a grid search for the optimal IPL parameter μ^ and the stripe size $\hat{s}_\ell$. With reference to Fig. 3, this corresponds to selecting the proper red curve (controlled by the μ parameter) to match the blue curve (empirical CCDF $F_{emph}(\tau)$) which is controlled by the stripe size selection. An alternative and more computationally effective approach is to utilize iterative techniques such as gradient descent or the Newton-Raphson method to iteratively determine a local minimum of the KS statistic in Eq. 5.

To demonstrate the effectiveness of the proposed KS-based stripe size and μ estimator, we evaluate its bias and variance using synthetic duration times generated via the Mittag-Leffler map, see, e.g., Huillet (2016). Fig. 4 shows the mean (blue) and variance (red), averaged over 100 Monte Carlo independent trials of the KS-based (solid curves) and MDEA-based μ estimator (dashed-curves) versus the number of data samples $N_{samp}$. It can be seen that as $N_{samp}$ increases, both the bias and variance of the KS-based estimator decrease; the same is also true for the MDEA estimator utilized to estimate the complexity scale δ and subsequently transforming it into μ when the KS-based stripe size estimate is used as input. However, this is not the case for MDEA when the stripe size is not correctly selected, in which case both the estimated bias and variance deviate as $N_{samp}$ increases.

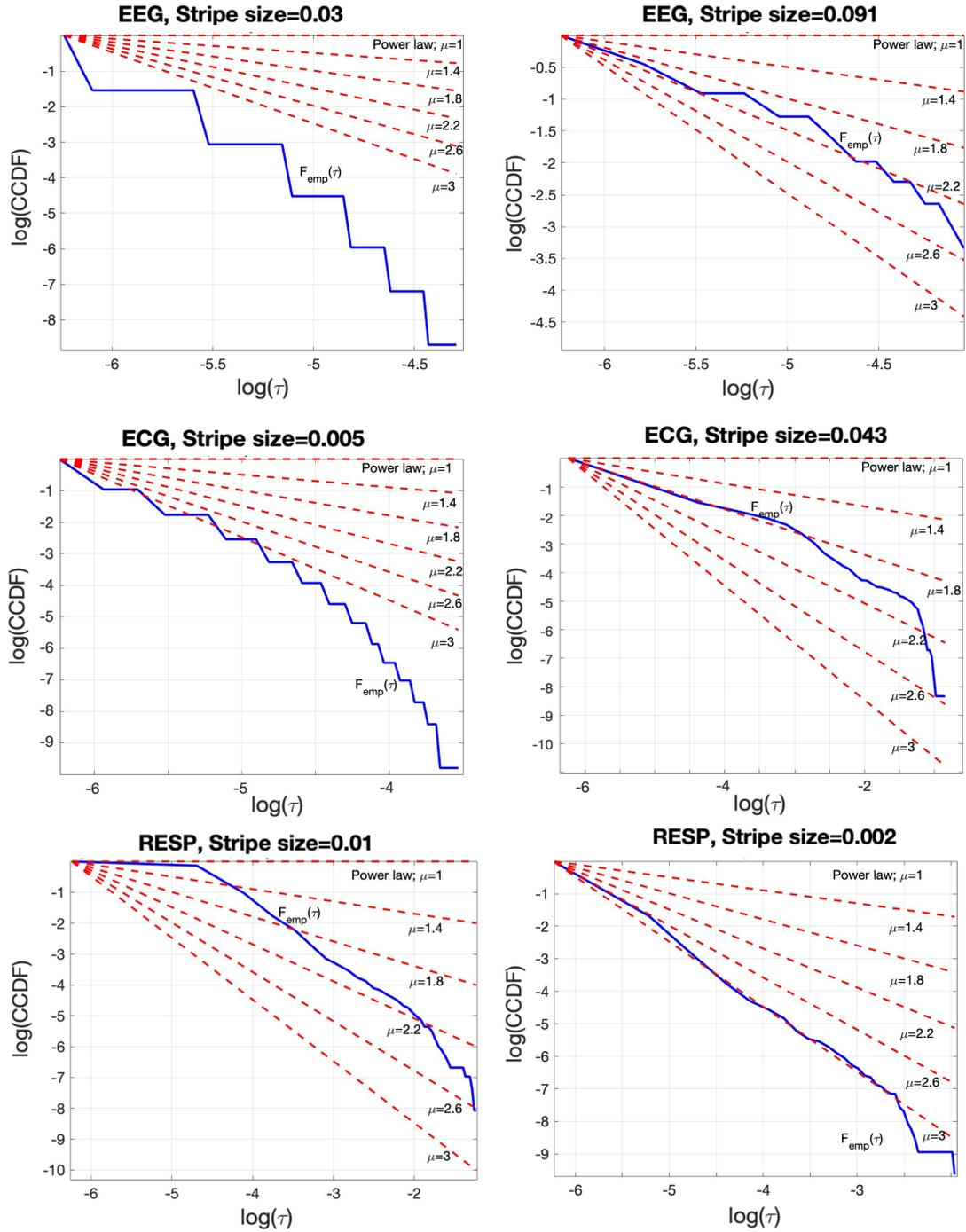

Figure 3: The impact of stripe size selection in the CE waiting-time PDF. Left column shows the no-IPL shape of the empirical waiting times PDF (blue curve) when the stripe size is not properly selected. Clearly, the blue curve corresponding to $F_{emp}(\tau)$ [cf. Eq. 2] does not fit any of the theoretical IPL PDFs indicated by the red curves for values of μ starting at μ = 1 and increasing by 0.4 as the red curves approach the horizontal axis. The right column indicates the IPL nature of the empirical CCDF $F_{emp}(\tau)$ defined in Eq. 2 (blue curve) when the stripe size is properly selected by minimizing the KS statistic to fit an IPL PDF.

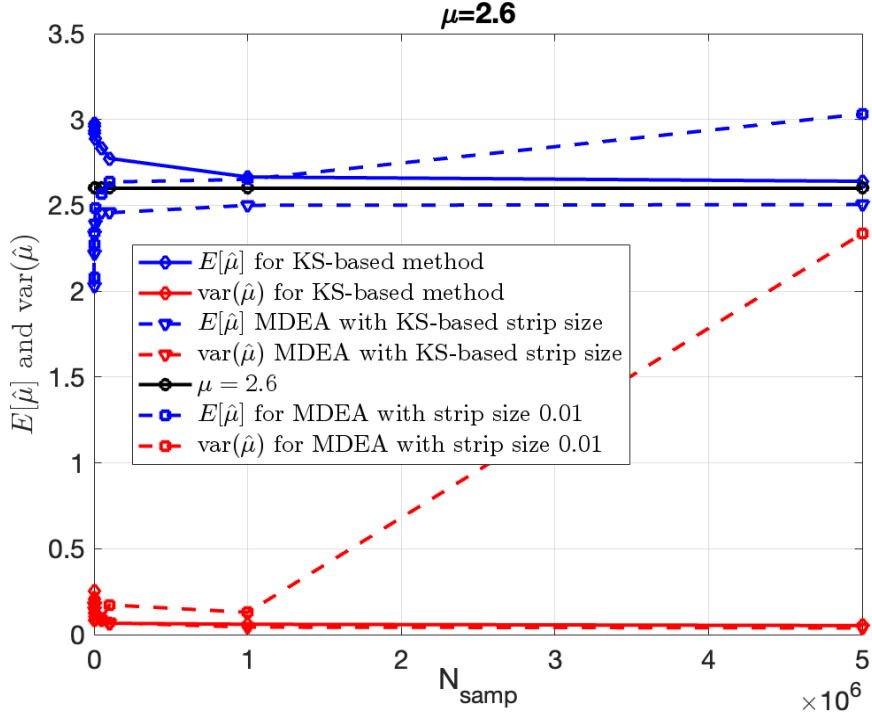

Figure 4: Bias and variance of KS-based and MDEA-based estimation of IPL parameter μ versus number of samples $N_{samp}$ utilized for the estimation.

Fig. 5, using synthetic duration times derived from the Mittag-Leffler map, shows the strength of KS-based stripe size selection to track changes in the IPL parameter μ. The nonlinear change in μ depicted by the black arrow in Fig. 5 essentially serves as a synthetic setting emulating real physiological signals whose complexity is constantly changing across time. This figure shows that the proposed KS-based method (corresponding μ estimates are visualized by the red and magenta trajectories in Fig. 5) tracks the nonlinear jump in IPL parameter μ. Further, MDEA with a fixed stripe size value does not track as effectively the time-varying μ.

We estimated stripe sizes using the stripe-size search function Stripe_size_search.m in Github (2024) across the entire data set (obtained during a Go-Nogo shooting simulation) consisting of 27 subjects (3 removed due to excessive EEG artifacts) in each of two-task conditions (low and high time stress) in 30 *sec* sliding windows with 20 *sec* overlap (N = 190, 938; total number of time windows).

Figure 6 shows the distributions of stripe sizes obtained for each EEG channel and the ECG and RESP ONs (raw and filtered). Note the relatively narrow distribution of stripe sizes for the EEG and ECG data but sparse and highly variable stripe sizes for the raw RESP data. We chose to use the median stripe size derived for each subject (i.e., individualized parameter estimates), which worked well for EEG and ECG but not for the RESP data. Table 2 summarizes descriptive statistics of stripe size estimates over all subjects, task conditions, and moving windows.

For RESP data, we encounter a different issue affecting the MDEA analysis, namely its highly periodic, deterministic characteristics. In extending our analyses to subjects beyond our original work (Mahmoodi et al. (2023a)), we discovered contradictory results. Originally, we observed significant scaling synchronization among EEG, ECG, and RESP. However, in subsequent analyses while testing the generalizability of these findings to other subjects in the experiment, we observed that the scaling of the RESP time series does not systematically synchronize with the EEG and ECG scaling when stripe size and linear fit parameters are chosen algorithmically as described above. The highly oscillatory nature of the RESP time series consists of more deterministic characteristics and may pose an issue for MDEA as events detected are not randomly distributed, which is an assumption of the theory.

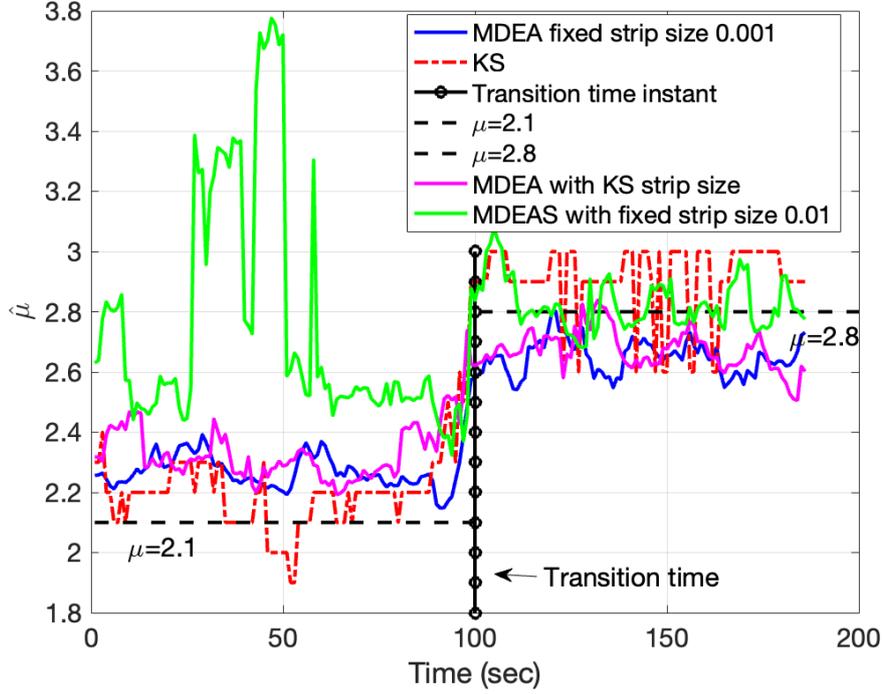

Figure 5: Tracking ability of IPL parameter μ using MDEA with automated stripe size selection by minimizing the KS statistic.

|          | Mean  | Med   | STD   | Max    | Min  | Range  |
|----------|-------|-------|-------|--------|------|--------|
| EEG      | 10.9  | 10.3  | 3.8   | 333.3  | 10.0 | 323.3  |
| ECG      | 35.4  | 33.3  | 16.9  | 333.3  | 10.0 | 323.3  |
| RESP     | 415.6 | 500.0 | 283.6 | 1000.0 | 10.0 | 990.0  |
| RESP Filt| 169.6 | 146.6 | 129.8 | 1402.7 | 10.0 | 1392.7 |

Table 2: Descriptive statistics of stripe size estimations across 190, 938 distinct 30 *sec* moving windows.

To address this situation, we reasoned that removing the highly periodic component of the RESP time series may improve the analysis. Consequently, we high-pass filtered the RESP signal at 2 Hz to minimize the peaks in the spectrum between approximately 0.25 and 0.75 Hz) (see Fig. 7). The filter used was a Kaiser high-pass filter of order 8192 whose frequency response is almost ideal, i.e., flat frequency response in the higher frequency range. The almost ideal high-pass behavior of the filter used allows the IPL spectra to be preserved above the filtered frequencies while suppressing the non-IPL components in the lower frequencies. Analyzing the filtered RESP time series improved the analysis since the IPL spectra was preserved (see Figs. 9 and 10 in the Results section), and the distributions of stripe sizes was more narrowly distributed (see Fig. 6). Figure 8 shows a comparison of the $S(w)$ vs. $log(w)$ plots for unfiltered (lower left) and filtered (lower right) RESP data. A word of caution is in order here, that is, we ask the question what does it mean to remove the dominant feature of the RESP time series? Although doing so yields improved CS results, we need to delve further into why this is the case and what are the implications with respect to the interpretability of the analyses. As another alternative, we also differentiated the raw RESP time series to remove the low-frequency periodicity, while preserving the velocity of changes in the time series prior to MDEA analyses and found comparable results to filtering in terms of scaling exponents and CS with EEG and ECG (results not shown). We are currently investigating this issue further, both theoretically and analytically.

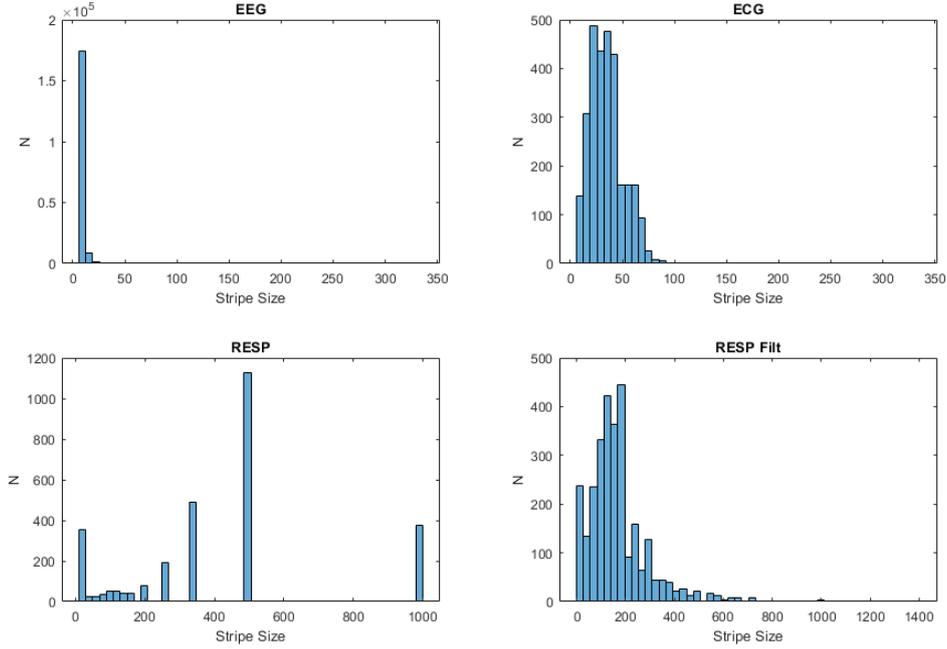

Figure 6: Distributions of stripe sizes determined across 30 *sec* moving windows for EEG, ECG, RESP, and filtered RESP across 27 subjects in low and high time stress conditions.

*2.3.2. Automated fit parameter region for estimating* δ *- scaling*

To address the issue of automatically determining the linear fit region of the $S(w)$ vs. $log(w)$ plot generated by MDEA, we use the ischange() function in Matlab in which the 'linear' method is designed to detect discontinuities, see function findLinearPortion_v2.m and findTwoLinearPortions.m in Github (2024). These points represent locations where the linear relationship shifts or breaks, indicating transitions between distinct linear segments. If detected, the indices of these change points are stored for further analysis, with each set of indices examined to determine if it meets the criteria for a valid scaling index.

A critical aspect is the adaptive adjustment of the threshold parameter, which originates from the ischange() function and determines the sensitivity of detecting these discontinuities. The adjustment is done in small increments to account for data variability and aims to detect the longest linear region that complies with and exceeds the minimum length requirement to calculate valid scaling indices. The $S(w)$ vs. $log(w)$ plot for EEG data is typically linear, as found in this particular study. Therefore, due to the power-law relationship, only one change point is often found. This change point signifies where the characteristics of the distribution shift, leading to the observed drop-off from the initial linear trend, highlighting the presence of fewer large values than expected in a typical power law, and defining the start of a heavy tail. When this occurs, but still follows an overarching linear pattern, the first change point is set at 20% of the entropy plot, with the next change point typically occurring at the beginning of the heavy tail.

In some instances, distortions can manifest as nonlinear behavior at the onset of the $S(w)$ vs. $log(w)$ plot, where the distribution might not follow the expected straight line, due to data noise or artifacts that can cause inconsistencies in measurements, particularly at low values of the $log(w)$ plot. When this occurs, the ischange() function detects two points autonomously and fits the linear region to this now intermediate section to determine the δ scaling index, see function findTwoLinearPortions.m in Github (2024).

For ECG data, this methodology changes slightly. ECG data, characterized by its repetitive pattern, affects the linearity of entropy calculations, resulting in the identification of multiple scaling indices (short, intermediate, and long time scales). Specifically, in linear cases, there is typically one scaling index, while in nonlinear cases, two or three scaling indices are identified due to short-, intermediate-, and long-term correlations between the events. In Fig.

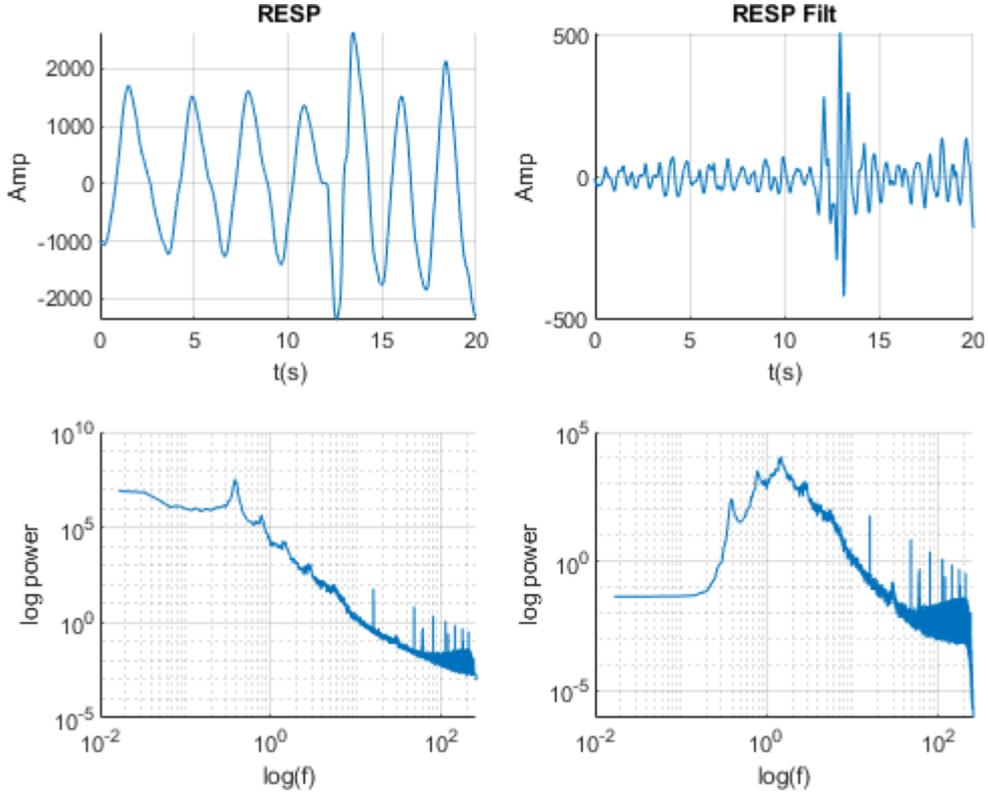

Figure 7: RESP time series zoomed to 20 *sec* (upper) and spectra (lower) unfiltered (left) and high-pass filtered (right) over entire 675 *sec* task period.

8 (top right) $S(w)$ vs. $log(w)$ plot for ECG reveals three clear segments, contrasting with the relatively linear EEG patterns. To create a holistic picture of complexity, this method focuses on identifying and tracking the middle linear segment of the ECG entropy to assess the scaling indices. This can be easily modified to characterize the scaling behavior of the time series over different ranges of fluctuations.

Just as with the stripe size selection parameter, the linear fit parameters are also determined as the median start and end points for the regression across all moving windows, so that all estimates for a given subject are constant across windows, although may be different across data types (EEG, ECG, RESP) within each subject.

## 3. Results

In Fig. 9 we plot the complexity scaling indices $\delta$s, returned by MDEA using the KS-based estimated stripe size values combined with the linear fit method described above, of 30 *sec* sliding windows of data with 20 *sec* overlap for an example subject (i.e., using individualized parameter estimates for that subject). The figure illustrates strong synchronization of scaling indices between EEG and ECG, but not with RESP (see Table 3). Consequently, we high-pass filtered the RESP above 2 Hz to minimize the deterministic oscillatory component of the time series. This approach revealed a significant improvement in synchronization of RESP with EEG and ECG as can be seen in Fig. 10 (see Table 4). Thus, KS-based stripe size selection combined with MDEA clearly can uncover CS patterns in the three heterogeneous time series advocating the effectiveness of automated KS-based stripe size estimation, given appropriate data considerations.

Across all 54 datasets (27 subjects in each of 2 conditions), we found significant CS among EEG, ECG, and filtered Resp in 24 datasets (each pairwise correlation $p < .05$) and lack of significance in 30 datasets. For significant

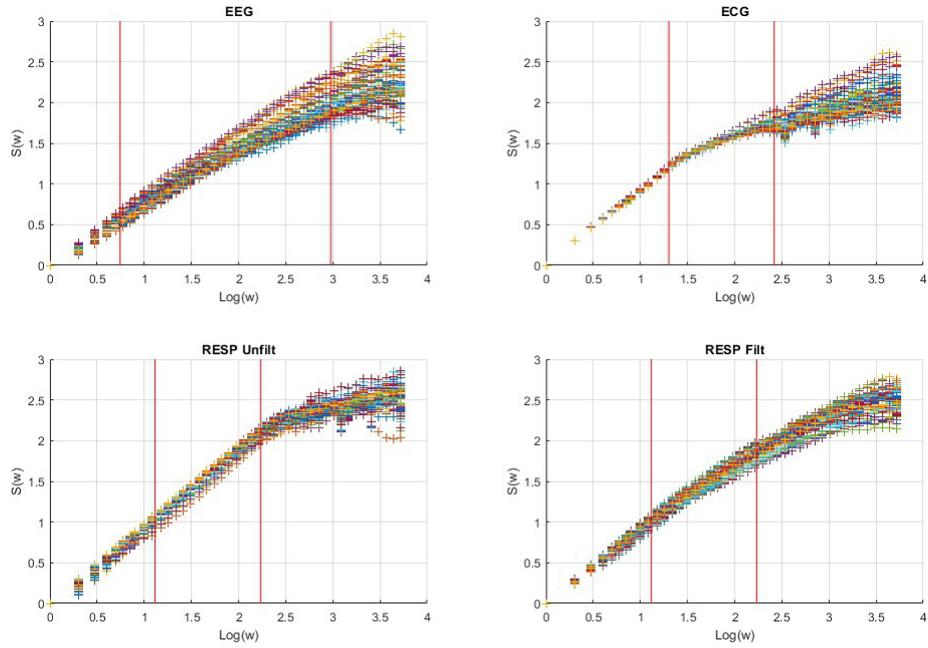

Figure 8: Entropy ($S(w)$) vs. log window length ($log(w)$) over sliding windows for EEG, ECG, and unfiltered and filtered RESP. Red lines demarcate the median fit regions applied for estimating delta scaling indices.

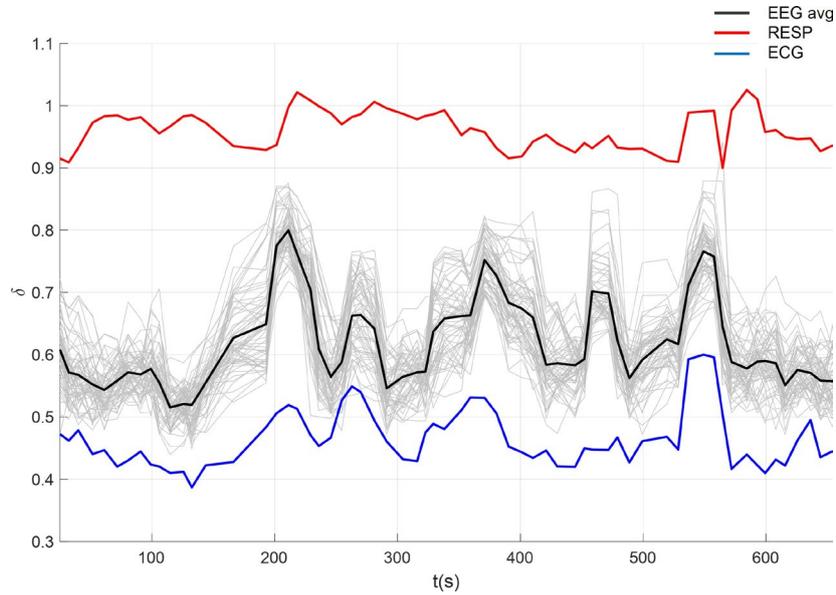

Figure 9: Complexity synchronization among EEG, ECG, and RESP using automated parameters selection.

CS, we found 11 in the Low and 13 in the High time stress condition. Further, 17 of 27 unique subjects exhibited significant CS in at least one condition, while 7 subjects exhibited significant CS in both Low and High time stress conditions.

|      | EEG  | ECG | RESP |
|------|------|-----|------|
| EEG  | -    |     |      |
| ECG  | .74* | -   |      |
| RESP | .06  | .20 | -    |

Table 3: Correlations among scaling indices of the mean of 64 EEG channels, ECG, and unfiltered RESP (*p< .01).

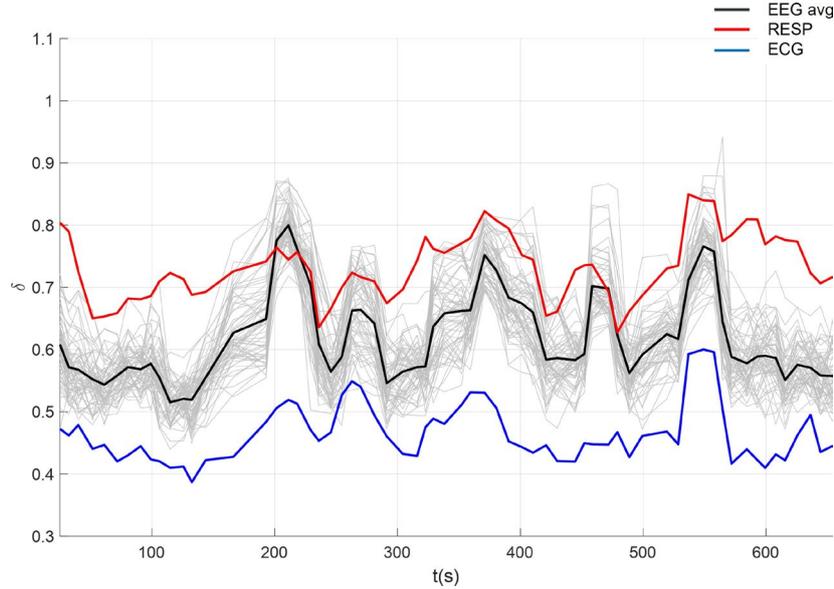

Figure 10: CS among EEG, ECG, and filtered RESP using automated parameters selection following high-pass filtering of the RESP time series.

|      | EEG  | ECG  | RESP |
|------|------|------|------|
| EEG  | -    |      |      |
| ECG  | .74* | -    | -    |
| RESP | .59* | .54* | -    |

Table 4: Correlations among scaling indices of the mean of 64 EEG channels, ECG, and high-pass filtered RESP (*p< .01).

For significant CS between the EEG and ECG, we found 46 of 54 datasets (23 in each Low and High time stress conditions) and only 8 which lacked significant CS between EEG and ECG (4 in Low and 4 in High time stress conditions). Further, 26 of 27 unique subjects exhibited significant CS in at least one condition, while 20 subjects exhibited significant CS in both Low and High time stress conditions.

For the CS between the EEG and filtered RESP, we found 32 of 54 datasets (17 in Low and 15 in High time stress conditions) with significant and 22 that lacked significant CS between EEG and filtered RESP. Further, 20 of 27 unique subjects exhibited significant CS in at least one condition, while 12 subjects exhibited significant CS in both Low and High time stress conditions.

For significant CS between the ECG and filtered RESP, we found 31 of 54 datasets (14 in Low and 17 in High time stress conditions) and 23 which lacked significant CS between ECG and filtered RESP. Further, 21 of 27 unique subjects exhibited significant CS in at least one condition, while 10 subjects exhibited significant CS in both Low and High time stress conditions.

Overall, these results indicate relatively higher CS between EEG and ECG (brain-heart coupling) than either EEG and RESP or ECG and RESP. In the Supplementary Materials, we provide representative examples from three additional subjects in each low and high time stress conditions. We plan to further investigate these differences across subjects and tasks to determine whether task performance is related to these outcomes or perhaps some other behav-ior (e.g., movement kinematics). Further, we are investigating the extent to which signal quality, pre-processing, or various decomposition approaches affect these results. We also intend to further investigate CS among all pairwise channels (64 EEG, ECG, and RESP) to determine whether certain channels or sub-networks of EEG deltas are specifically coupling more strongly amongst EEG channels, as well as with the ECG and RESP, beyond just the average of all EEG channels. We also intend to investigate other measures of coupling in CS beyond correlation, e.g., whether distance or topological differences influence CS within the brain.

## 4. Discussion and Conclusions

The fractal nature of heterogeneous neurophysiological time series suggests the lack of any one frequency or scale dominating the dynamics of any physiologic process West (2006). Therefore, holistic theories and methods invoking multifractal dimensionality of vastly different neurophysiological and behavioral processes interacting in nonlinear dynamic ways offer new promising alternatives for better understanding communication among NoONs (complex systems). Herein we have attempted to advance the state of the art in objectifying and automating parameter selection for MDEA and its application to CS analysis. Two here-to-fore outstanding issues have been addressed and advanced in this work. objectively determining: 1) the stripe sizes and 2) the linear fit regions of the different ON time series in MDEA. This progress facilitates both our research and that of others in replicating and further testing as well as testing the theories and methods presented herein and enables the analyses to be conducted on large-scale datasets.

Communication among NoONs coexist via several forms of coupling simultaneously Bartsch and Ivanov (2014); Bartsch et al. (2015); Ivanov (2021). The form of coupling observed through CS is a new phenomenon which requires further advances in theory, modeling, and empirical research-analyses. Comparisons with other theories and methods is also needed (e.g., see Table 1) to better understand the principles and mechanisms through which heterogeneous but integrated ONs within NoONs interact to optimize human health and performance.

The nonlinear mutual interactions between human ONs and NoONs give rise to complex dynamics operating by the information gradient among them. This rather benign observation is, in fact, a profound result in that it is a statement of the physiologic system being driven by an information force and not a mechanical force. Social organization and physiological function are both driven by dynamic interactions among complex ONs, where ONs can mean organ-networks or organization-networks. In both contexts what is of importance is the manner in which information is shuttled back and forth between such non-physical networks and whether there exists a general principle that guides that flow of information in the same way that energy flow determines forces in physical networks. Such a principle has been identified and is discussed in a number of places, see e.g., West (2016). One consequence of the existence of this principle is a new kind of force; a force based on the relative complexity of the interacting networks producing an information gradient. This information force reduces to the entropic force in physical networks but in non-physical ONs results from gradients in the complexity of the phenomenon being studied. We think that this novel method will enable the study of the brain's self-organization in real-time.

*4.1. Future Research*

Findings observed herein need to be generalized to additional subjects in the experiment leveraged here, as well as to data from other diverse datasets, including those featuring simultaneously recorded time series and point processes from neural, physiological, behavioral, environmental, social, and biological systems. New experiments must also be designed to more specifically test theories and hypotheses and address outstanding research questions.

As we outlined in the Introduction, several factors, including experiment design, subjects, task and conditions, data features and characteristics, signal processing and analysis approaches, and missing and artifact-contaminated data considerations need to be further systematically tested and validated within the CS analysis framework. We are currently testing the effects of various types and levels of EEG artifacts on MDEA scaling and CS analysis, as well as the effectiveness of various stages and levels of artifact reduction, including mitigation strategies. We are also studying CS among multi-modal data of diverse nature and types, such as heart rate variability, gate cycles, kinematic, kinetic, and metabolic measurements, neuromuscular and ocular activity, and several behavioral measures that may be continuous or intermittent to better understand how CS relates to changes in task performance and cognitive-affect-state changes. Future research is also needed to test some of the assumptions of CE theory, such as the independence of $\tau$s generated by MDEA analysis and whether crucial event rehabilitation therapy (CERT) can be used to enhance performance and health by driving CS among neurophysiological and behavioral NoONs West et al. (2023a). Although we present our theory and methods using neurophysiological data (EEG, ECG, and RESP), the approach should be generalizable across a diverse range of time series generated by complex systems transcending scientific disciplines from social network sciences to biosciences to engineering artificial intelligence.

**Conflict of Interest Statement**

The authors declare that the research was conducted in the absence of any commercial or financial relationships that could be construed as a potential conflict of interest.

**Author Contributions**

IS, SS, CJB, KM SEK developed and tested the parameter selection algorithms for MDEA; SEK, PJF, processed and analyzed neurophysiological data and applied CS analysis; PG, BJW, and KM led and guided the theoretical and analytical foundations underlying and motivating this research; SEK, IS, SS, KM, and BJW co-wrote the manuscript. DLB contributed program management, funding, and guidance which enabled this research. All authors critically assessed and discussed the results, and revised and approved the manuscript.

**Funding**

This research was sponsored by the Army Research Laboratory and was accomplished under Cooperative Agreement Number W911NF-23-2-0162. The views and conclusions contained in this document are those of the authors and should not be interpreted as representing the official policies; either expressed or implied, of the Army Research Laboratory or the U.S. Government. The U.S. Government is authorized to reproduce and distribute reprints for Government purposes notwithstanding any copyright notation herein.

**Acknowledgments**

We acknowledge Herbert F. Jelinek, Sara A. Nasrat, Yawer Shah, Hamid Khajoei, and Jace Singh for their contributions in conceptual discussions, algorithm testing and development, and editorial comments on this manuscript.

**Data Availability Statement**

A sample dataset used for this study is contained in the file S037.SH.L.sess6.4chans.mat in the following Github link: https://github.com/Yanniml/Kolmogorov-Smirnov-MDEA

**Code Availability Statement**

The codes used in this study are available in the following GitHub link: https://github.com/Yanniml/Kolmogorov-Smirnov-MDEA

Individual functions for the methods described in this paper are provided. Further, script MDEAoptstripe.m summarizes the order and the way these methods are put together to process the physiological signals.

# References

Allegrini, P., Menicucci, D., Bedini, R., Fronzoni, L., Gemignani, A., Grigolini, P., West, B.J., Paradisi, P., 2009. Spontaneous brain activity as a source of ideal 1/f noise. Physical Review E 80, 061914.

Alskafi, F.A., Khandoker, A.H., Marzbanrad, F., Jelinek, H.F., 2023. Eeg-based emotion recognition using sub-band time-delay correlations, in: 2023 45th Annual International Conference of the IEEE Engineering in Medicine & Biology Society (EMBC), IEEE. pp. 1–4.

Bartsch, R.P., Ivanov, P.C., 2014. Coexisting forms of coupling and phase-transitions in physiological networks, in: Nonlinear Dynamics of Electronic Systems: 22nd International Conference, NDES 2014, Albena, Bulgaria, July 4-6, 2014. Proceedings 22, Springer. pp. 270–287.

Bartsch, R.P., Liu, K.K., Bashan, A., Ivanov, P.C., 2015. Network physiology: how organ systems dynamically interact. PloS one 10, e0142143.

Buongiorno Nardelli, M., Culbreth, G., Fuentes, M., 2022. Towards a measure of harmonic complexity in western classical music. Advances in Complex Systems 25, 2240008.

Canolty, R.T., Knight, R.T., 2010. The functional role of cross-frequency coupling. Trends in cognitive sciences 14, 506–515.

Catrambone, V., Barbieri, R., Wendt, H., Abry, P., Valenza, G., 2021. Functional brain–heart interplay extends to the multifractal domain. Philosophical Transactions of the Royal Society A 379, 20200260.

Corder, G.W., Foreman, D.I., 2014. Nonparametric Statistics: A Step-by-Step Approach. Wiley, The city.

Costa, M., Goldberger, A.L., Peng, C.K., 2005. Multiscale entropy analysis of biological signals. Physical Review E—Statistical, Nonlinear, and Soft Matter Physics 71, 021906.

Culbreth, G., 2021. Information and Self-Organization in Complex Networks. Ph.D. thesis. University of North Texas.

Deco, G., Vidaurre, D., Kringelbach, M., 2021. Revisiting the global workspace orchestrating the hierarchical organization of the human brain. nature human behaviour, 5 (4), 497–511.

Faes, L., Marinazzo, D., Stramaglia, S., 2017. Multiscale information decomposition: Exact computation for multivariate gaussian processes. Entropy 19, 408.

Gao, J., Hu, J., Liu, F., Cao, Y., 2015. Multiscale entropy analysis of biological signals: a fundamental bi-scaling law. Frontiers in computational neuroscience 9, 64.

Github, 2024. https://github.com/yanniml/kolmogorov-smirnov-mdea. URL: https://github.com/Yanniml/Kolmogorov-Smirnov-MDEA.

Herrero, J.L., Khuvis, S., Yeagle, E., Cerf, M., Mehta, A.D., 2018. Breathing above the brain stem: volitional control and attentional modulation in humans. Journal of neurophysiology .

Huillet, T.E., 2016. On mittag-leffler distributions and related stochastic processes. Journal of Computational and Applied Mathematics 296, 181–211. doi:https://doi.org/10.1016/j.cam.2015.09.031.

Ivanov, P.C., 2021. The new field of network physiology: building the human physiolome.

Jelinek, H.F., Tuladhar, R., Culbreth, G., Bohara, G., Cornforth, D., West, B.J., Grigolini, P., 2021. Diffusion entropy vs. multiscale and renyi entropy to detect progression of autonomic neuropathy. Frontiers in Physiology 11, 607324.

Kerick, S.E., Asbee, J., Spangler, D.P., Brooks, J.B., Garcia, J.O., Parsons, T.D., Bannerjee, N., Robucci, R., 2023. Neural and behavioral adaptations to frontal theta neurofeedback training: A proof of concept study. Plos one 18, e0283418.

Kotiuchyi, I., Pernice, R., Popov, A., Kharytonov, V., Faes, L., 2021. Mutual information analysis of brain-heart interactions in epileptic children, in: 2021 Signal Processing Symposium (SPSympo), IEEE. pp. 133–137.

Kumar, M., Singh, D., Deepak, K., 2020. Identifying heart-brain interactions during internally and externally operative attention using conditional entropy. Biomedical Signal Processing and Control 57, 101826.

Leistritz, L., Pester, B., Doering, A., Schiecke, K., Babiloni, F., Astolfi, L., Witte, H., 2013. Time-variant partial directed coherence for analysing connectivity: a methodological study. Philosophical Transactions of the Royal Society A: Mathematical, Physical and Engineering Sciences 371, 20110616.

Li, P., Li, K., Liu, C., Zheng, D., Li, Z.M., Liu, C., 2016. Detection of coupling in short physiological series by a joint distribution entropy method. IEEE Transactions on Biomedical Engineering 63, 2231–2242.

Lin, A., Liu, K.K., Bartsch, R.P., Ivanov, P.C., 2016. Delay-correlation landscape reveals characteristic time delays of brain rhythms and heart interactions. Philosophical Transactions of the Royal Society A: Mathematical, Physical and Engineering Sciences 374, 20150182.

Liu, K.K., Bartsch, R.P., Ma, Q.D., Ivanov, P.C., 2015. Major component analysis of dynamic networks of physiologic organ interactions, in: Journal of Physics: Conference Series, IOP Publishing. p. 012013.

Lobier, M., Siebenhühner, F., Palva, S., Palva, J.M., 2014. Phase transfer entropy: a novel phase-based measure for directed connectivity in networks coupled by oscillatory interactions. Neuroimage 85, 853–872.

Mahmoodi, K., Kerick, S.E., Grigolini, P., Franaszczuk, P.J., West, B.J., 2023a. Complexity synchronization: a measure of interaction between the brain, heart and lungs. Scientific Reports 13, 11433.

Mahmoodi, K., Kerick, S.E., Grigolini, P., Franaszczuk, P.J., West, B.J., 2023b. Temporal complexity measure of reaction time series: Operational versus event time. Brain and Behavior 2023, e3069.

Martin, A., Guerrero-Mora, G., Dorantes-Méndez, G., Alba, A., Méndez, M.O., Chouvarda, I., 2015. Non-linear analysis of eeg and hrv signals during sleep, in: 2015 37th Annual International Conference of the IEEE Engineering in Medicine and Biology Society (EMBC), IEEE. pp. 4174–4177.

Marwan, N., Romano, M.C., Thiel, M., Kurths, J., 2007. Recurrence plots for the analysis of complex systems. Physics reports 438, 237–329.

Marzbanrad, F., Yaghmaie, N., Jelinek, H.F., 2020. A framework to quantify controlled directed interactions in network physiology applied to cognitive function assessment. Scientific Reports 10, 18505.

Nyquist, H., 1928. Certain topics in telegraph transmission theory. Transactions of the American Institute of Electrical Engineers 47, 617–644.

Pan, S., Han, T., Tan, A.C., Lin, T.R., 2016. Fault diagnosis system of induction motors based on multiscale entropy and support vector machine with mutual information algorithm. Shock and Vibration 2016, 5836717.

Pernice, R., Zanetti, M., Nollo, G., De Cecco, M., Busacca, A., Faes, L., 2019. Mutual information analysis of brain-body interactions during different levels of mental stress, in: 2019 41st Annual International Conference of the IEEE Engineering in Medicine and Biology Society (EMBC), IEEE. pp. 6176–6179.

Petzschner, F.H., Weber, L.A., Wellstein, K.V., Paolini, G., Do, C.T., Stephan, K.E., 2019. Focus of attention modulates the heartbeat evoked potential. NeuroImage 186, 595–606.
Piper, D., Schiecke, K., Pester, B., Benninger, F., Feucht, M., Witte, H., 2014. Time-variant coherence between heart rate variability and eeg activity in epileptic patients: an advanced coupling analysis between physiological networks. New Journal of Physics 16, 115012.
Reshef, D.N., Reshef, Y.A., Finucane, H.K., Grossman, S.R., McVean, G., Turnbaugh, P.J., Lander, E.S., Mitzenmacher, M., Sabeti, P.C., 2011. Detecting novel associations in large data sets. science 334, 1518–1524.
Rosenblum, M.G., Pikovsky, A.S., Kurths, J., 1996. Phase synchronization of chaotic oscillators. Physical review letters 76, 1804.
Scafetta, N., Grigolini, P., 2002. Scaling detection in time series: Diffusion entropy analysis. Physical Review E 66, 036130.
Schandry, R., Sparrer, B., Weitkunat, R., 1986. From the heart to the brain: a study of heartbeat contingent scalp potentials. International Journal of Neuroscience 30, 261–275.
Schiecke, K., Schumann, A., Benninger, F., Feucht, M., Baer, K.J., Schlattmann, P., 2019. Brain–heart interactions considering complex physiological data: processing schemes for time-variant, frequency-dependent, topographical and statistical examination of directed interactions by convergent cross mapping. Physiological measurement 40, 114001.
Schreiber, T., 2000. Measuring information transfer. Physical review letters 85, 461.
Sugihara, G., May, R., Ye, H., Hsieh, C.h., Deyle, E., Fogarty, M., Munch, S., 2012. Detecting causality in complex ecosystems. science 338, 496–500.
Tort, A.B., Brankacˇk, J., Draguhn, A., 2018. Respiration-entrained brain rhythms are global but often overlooked. Trends in neurosciences 41, 186–197.
Tort, A.B., Komorowski, R., Eichenbaum, H., Kopell, N., 2010. Measuring phase-amplitude coupling between neuronal oscillations of different frequencies. Journal of neurophysiology 104, 1195–1210.
Vicente, R., Wibral, M., Lindner, M., Pipa, G., 2011. Transfer entropy—a model-free measure of effective connectivity for the neurosciences. Journal of computational neuroscience 30, 45–67.
West, B.J., 2006. Where medicine went wrong: Rediscovering the path to complexity. volume 11. World Scientific.
West, B.J., 2016. Information forces. Journal of Theoretical Computational Science 3, 144.
West, B.J., Grigolini, P., Bologna, M., 2023a. Crucial Event Rehabilitation Therapy. Springer.
West, B.J., Grigolini, P., Kerick, S.E., Franaszczuk, P.J., Mahmoodi, K., 2023b. Complexity synchronization of organ networks. Entropy 25, 1393.

# 1. Supplementary Data

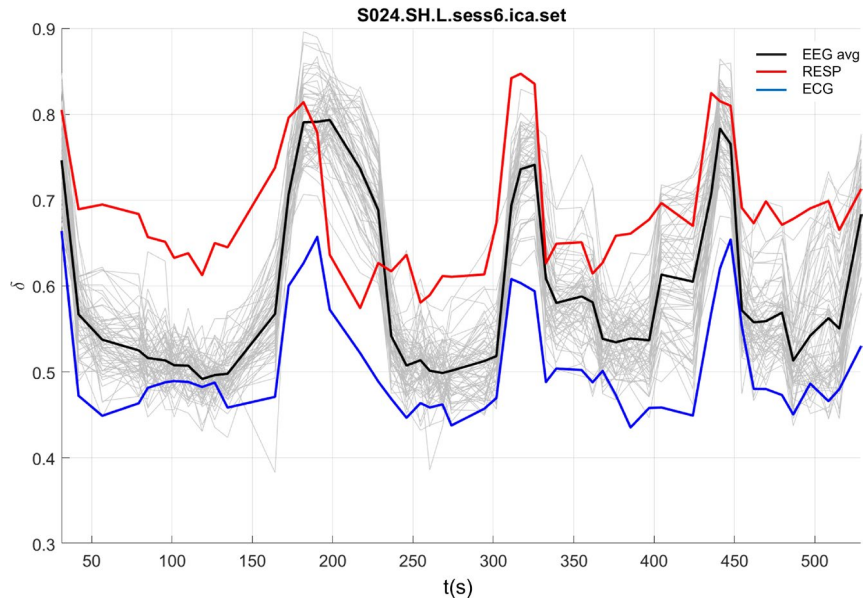

Figure 1: Significant complexity synchronization for subject S024 among EEG, ECG, and filtered RESP in Low time stress condition.

|      | EEG  | ECG  | RESP |
|------|------|------|------|
| EEG  | -    |      |      |
| ECG  | .87* | -    |      |
| RESP | .69* | .77* | -    |

Table 1: Correlations among scaling indices of the mean of 64 EEG channels, ECG, and filtered RESP for S024 Low time stress condition (*p< .05).

|      | EEG  | ECG  | RESP |
|------|------|------|------|
| EEG  | -    |      |      |
| ECG  | .80* | -    |      |
| RESP | .77* | .86* | -    |

Table 2: Correlations among scaling indices of the mean of 64 EEG channels, ECG, and filtered RESP for S024 High time stress condition (*p< .05).

|      | EEG  | ECG  | RESP |
|------|------|------|------|
| EEG  | -    |      |      |
| ECG  | .54* | -    |      |
| RESP | .32* | .21  | -    |

Table 3: Correlations among scaling indices of the mean of 64 EEG channels, ECG, and filtered RESP for S029 Low time stress condition (*p< .05).

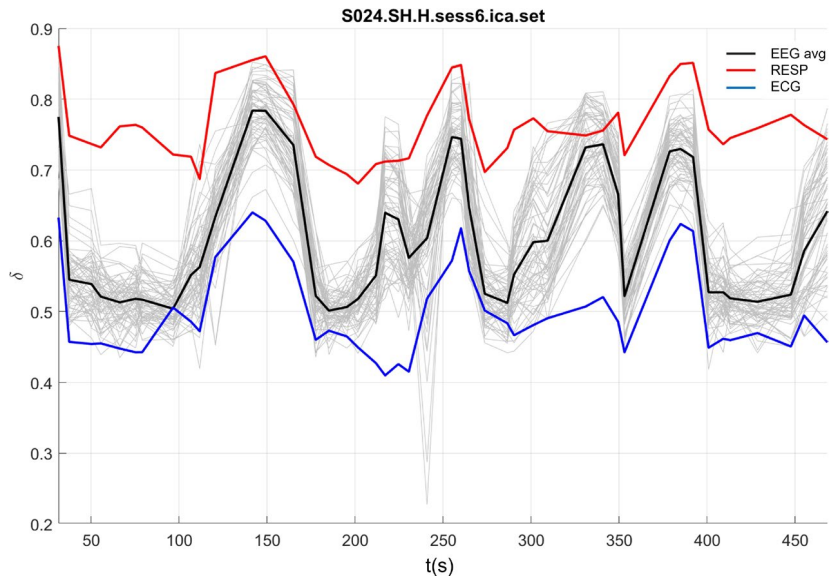

Figure 2: Significant complexity synchronization for subject S024 among EEG, ECG, and filtered RESP in High time stress condition.

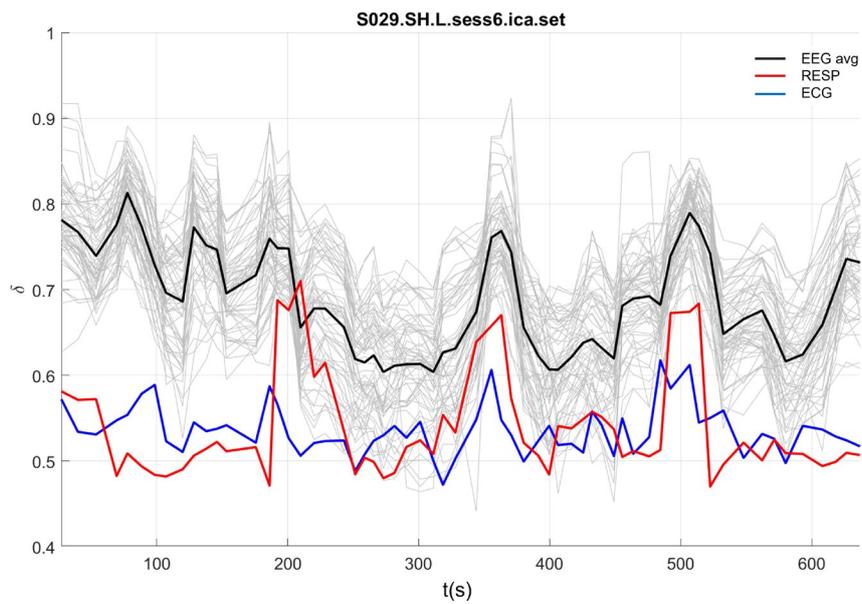

Figure 3: Non-significant complexity synchronization for subject S029 among EEG, ECG, and filtered RESP in Low time stress condition.

|      | EEG  | ECG  | RESP |
|------|------|------|------|
| EEG  | -    |      |      |
| ECG  | .37* | -    |      |
| RESP | .02  | .04  | -    |

Table 4: Correlations among scaling indices of the mean of 64 EEG channels, ECG, and filtered RESP for S029 High time stress condition (*p< .05).

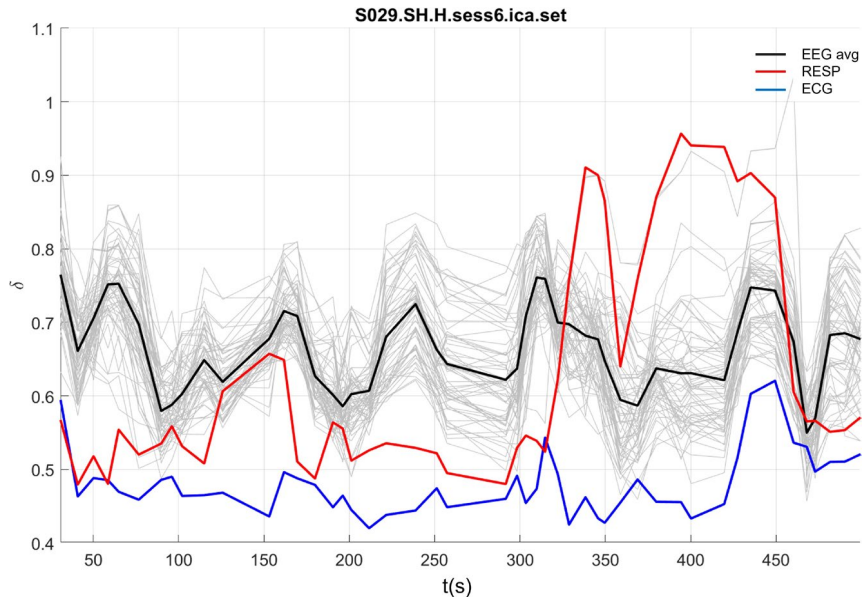

Figure 4: Non-significant complexity synchronization for subject S029 among EEG, ECG, and filtered RESP in High time stress condition.

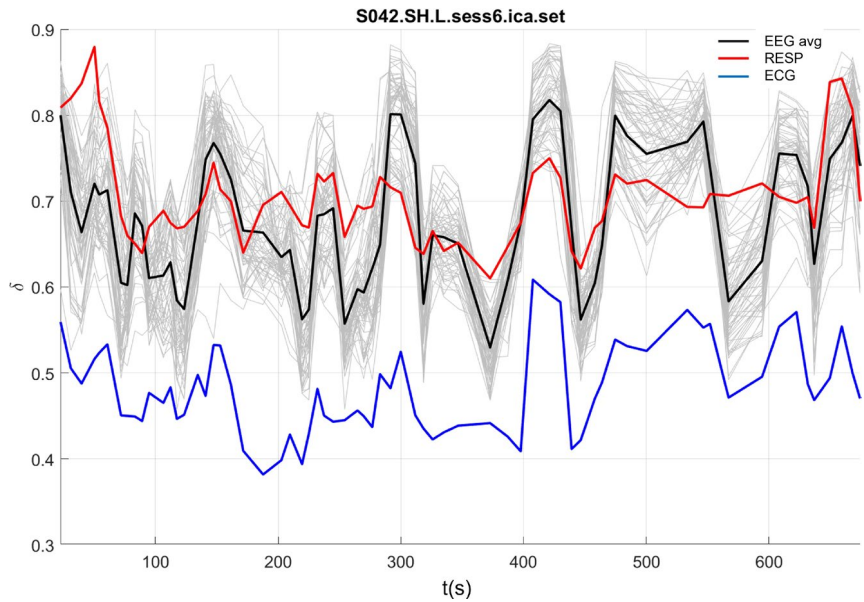

Figure 5: Significant complexity synchronization for subject S042 among EEG, ECG, and filtered RESP in Low time stress condition.

|      | EEG   | ECG   | RESP |
|------|-------|-------|------|
| EEG  | -     |       |      |
| ECG  | .74*  | -     |      |
| RESP | .53*  | .54*  | -    |

Table 5: Correlations among scaling indices of the mean of 64 EEG channels, ECG, and filtered RESP for S042 Low time stress condition (*p< .05).

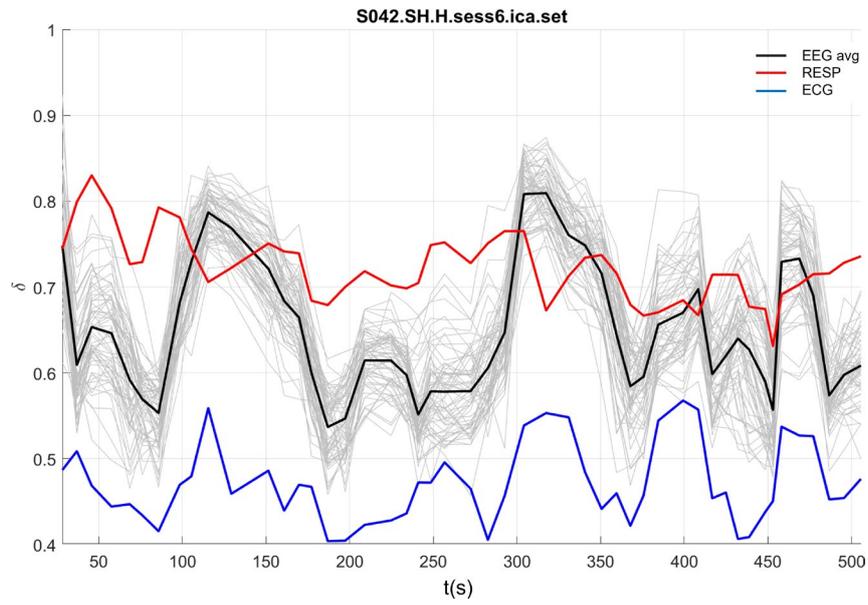

Figure 6: Non-significant complexity synchronization for subject S042 among EEG, ECG, and filtered RESP in High time stress condition.

|      | EEG  | ECG  | RESP |
|------|------|------|------|
| EEG  | -    |      |      |
| ECG  | .65* | -    |      |
| RESP | .10  | -.09 | -    |

Table 6: Correlations among scaling indices of the mean of 64 EEG channels, ECG, and filtered RESP for S042 High time stress condition (*p< .05).